# Energy-Efficient Cellular Communications Powered by Smart Grid Technology

Imtiaz Nasim, Mostafa Zaman Chowdhury, and Md. Syadus Sefat
Department of Electrical and Electronic Engineering
Khulna University of Engineering & Technology, Khulna-9203, Bangladesh
E-mail: imtiaz_nasim@hotmail.com, mzceee@yahoo.com, sefat_235711@yahoo.com

*Abstract*—The energy efficiency aspect of cellular networks is a vital topic of research over the recent days. As energy consumption is on the rise and the price of electricity is increasing very rapidly, necessity to reduce electricity usage in every aspect is becoming much more significant. The power grid infrastructure, from which the cellular networks attain the required electricity for operation is considering a significant change from the traditional electricity grid to the smart grid. The base stations, which are the main candidates for energy consumption in cellular networks, remain in operation even when there are a very few users or no user at all. In this paper, we propose an effective application scenario of the femtocell technology for power saving purpose of cellular networks. Effective deployment of femto-access-points in the required places can reduce the power usage of cellular networks by turning off the redundant base stations. Moreover, the signal to noise plus interference ratio is also enhanced in comparison to the traditional scheme. The base station turn off probability and the energy consumption in the overall network are analyzed. The comparison results show that the proposed energy efficient model can significantly reduce energy consumption and increase the service quality of the cellular users.

## I. INTRODUCTION

The smart grid *(SG)*, also called smart power grid, intelligent grid, intelligrid, future grid or intragrid, is an enhancement of the 20th century power grid. The traditional power grids are normally used to carry power in one way direction only. These power grids generally carry power from a few central generators to a large number of users or consumers. On the other hand, SG creates an automated and distributed advanced energy delivery network by using the two-way flow of electricity. The SG uses modern information technologies and is capable of distributing power in a more efficient way. It also emphasizes on energy saving of the different smart appliances and commodities [1].

Recently, rapidly rising energy price has led to a trend of studying the energy efficiency aspect of communication networks. The cellular operators pay a handsome amount of electricity bill which has become a significant portion of their total expenditure. The $CO_2$ emissions produced by wireless cellular networks all over the world are equivalent to those from more than 8 million cars. In a typical wireless cellular network, base stations (BSs) consumes $60 - 80\%$ energy of the overall cellular network's energy consumption. Almost 120,000 new base stations are deployed throughout the world every year to provide necessary support to the cellular users [2]. Our survey works suggest that a base station consumes more than 85% of its peak energy even when there is little or no traffic. Therefore, a dynamic operation of the cellular base stations based on the traffic is very much essential for the energy efficient approach of cellular networks. The proposed model in this paper is expected to reduce the energy consumption of cellular networks to a significant level. However, this model is much more applicable for some remote areas where there are a very few cellular users and a base station remains in operation continuously to serve those few users. Using the simulation results, we show that the proposed scheme can save energy of cellular networks and improve the quality of service (QoS) to a certain level.

The rest of this paper is organized as follows. Section II shows the proposed energy efficient model. Section III shows the performance analysis of the proposed scheme. Finally, conclusions are drawn in the last section.

## II. PROPOSED ENERGY EFFICIENT MODEL

Suppose a cellular network is powered by smart grid where there is a flow of electricity and communication simultaneously as shown in Fig. 1.

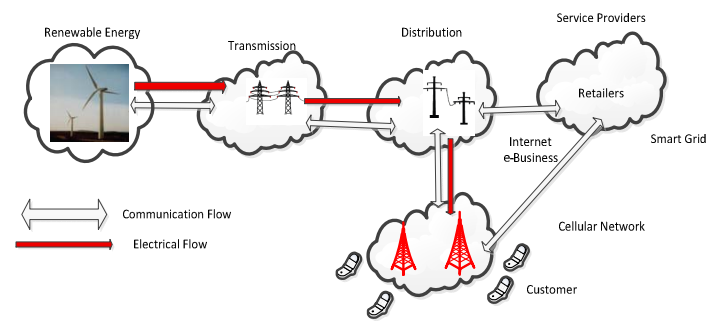

Figure 1. A cellular network powered by the smart grid [2].

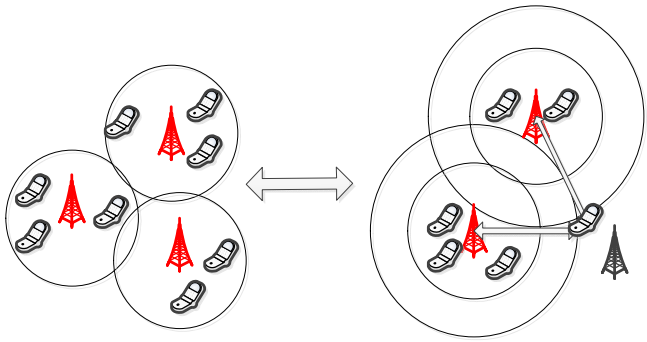

Figure 2. Coverage extension and energy saving with coordinated multipoint (CoMP) [2].

In [2], the authors proposed a model in which the redundant base station is turned off and the users of that network are served by coverage extension of the neighboring base stations using the coordinated multipoint method for energy saving purpose. This model is showed in Fig. 2.

If the redundant base station is turned off as shown in Fig. 2, a decent amount of energy can be saved if the existing user is served by the neighboring base stations. But we propose another model for saving energy of cellular networks using femtocell network which may turn out to be a very promising solution for saving more energy of cellular networks especially in remote areas.

In our proposed model, femto-access-points are installed in some places which have the most probability of carrying a mobile user in the entire cellular network served by a base station. The femto-access-points (FAPs) are low-power, small-size home-placed base stations that enhance the service quality for the indoor mobile users [3]. Fig. 3 shows an example of the proposed energy efficient model. When all the users remain inside the femtocellular coverage region, the macrocellular BS is allowed to turn off. But if any user stays outside the femtocellular coverage, the macrocellular BS must remain in operation to serve that particular user. Periodic checking of user's location and the QoS is tested by the dynamic checking application of the SG. Our purpose is not only saving energy of cellular networks but also to increase the service quality of the existing users. Femto-access-points are ideal for this case as these devices run with very low operating power and at the same time can enhance the QoS to a significant level.

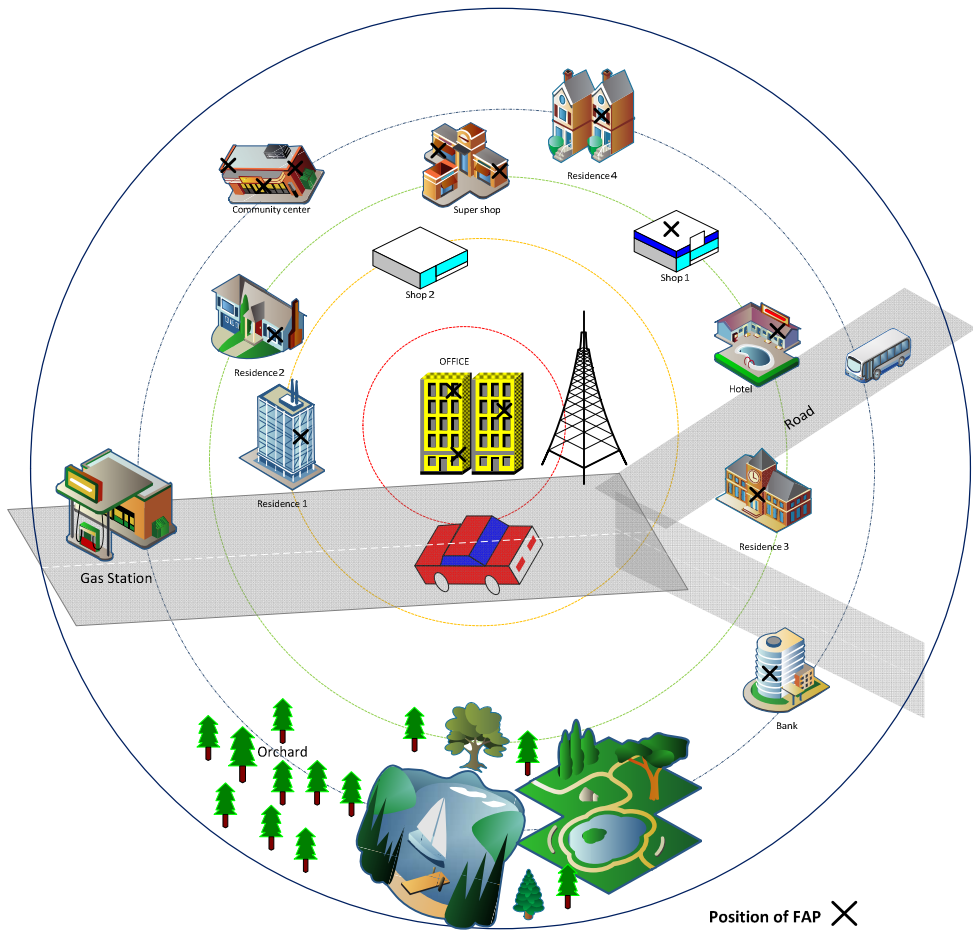

Figure 3. Proposed energy efficient model with femtocell/Macrocell integrated network.

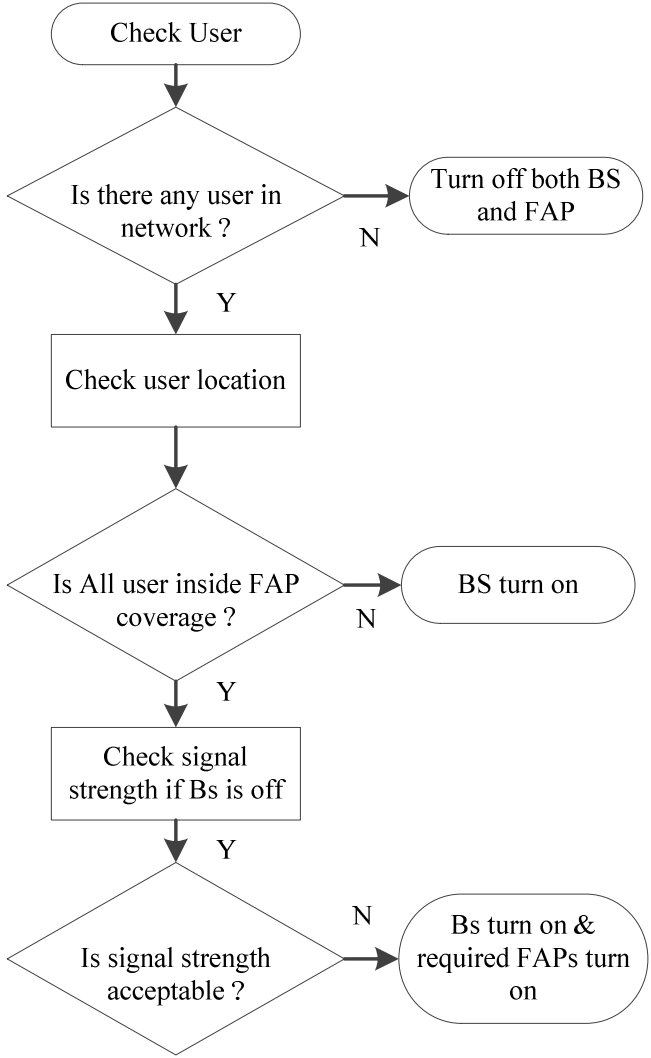

Figure 4. Flow chart showing the condition for the BS to be turned off in the proposed scheme.

Femtocells are small base stations with low transmission power, and with almost all cellular functionalities, but a limited coverage to end users in indoor environments. A femtocell allows service coverage extended to indoor environments, especially where wireless access is limited or unavailable, without expensive deployment cost.

A flow chart is shown in fig. 4 which shows the algorithms of the proposed scheme. The network will first check the existence of a user. If no user exists in the network, both the macrocellular BS and the FAPs will remain off. If any users exist, then the network will search for user's location. When any of the users remain outside the FAP coverage area, the BS will not be allowed to turn off. But, if all users exist inside the femtocellular network, then the network will check the signal strength available for the users if the BS is shut down. The BS will shut down if the signal strength is found within acceptable limit, otherwise; it will remain on in order to provide network support to the existing users. This model requires a periodic checking of the position of each user inside the network. By adopting this process, the macrocellular BS will sometimes remain turned on and sometimes remain turned off.

## III. PERFORMANCE ANALYSIS

This section provides the probability analysis performance and the signal quality performance of our proposed model.

### A. Probability Analysis

When a redundant BS is shut down and the users of that area are to be served using FAPs, then it is required to analyze the probability of the existence of a user under the femtocell coverage area. If all users are within the femtocell coverage area then the BS may remain shut down and service is provided to the user by the femtocell network, but if any user remains outside the femtocell coverage area, then the BS must be turned on in order to provide service to that user.

For simplicity in calculation, we have divided the total considered area into 4 sub-circles each with a fixed distance from the center point of the main circle. The coverage area of a macrocellular BS is assumed to be 500m. We performed our analysis assuming 15 FAPs in the total coverage area. All the stations considered in the coverage area have a specific factor according to the importance of that station for carrying users. The factors assumed for each station is given in Table I. The coverage area of a FAP is determined by

$$a = 4\pi r^2$$

where r is the radius of each FAP coverage.

If there exist stations with $n$ number of factors then Let, $A_1$ is the area under user existence probability factor $f_1$, $A_2$ is the area under user existence probability factor $f_2$, $A_3$ is the area under user existence probability factor $f_3$, and $A_n$ is the area under user existence probability factor $f_n$.

Let the remaining area where no FAPs are installed is $A_M$ with a user existence probability factor of $f_M$

The probability of a user to exist in the FAP coverage area is given by

$$P = \frac{f_1 \sum_{i=1}^{N_1} A_{1i} + f_2 \sum_{i=1}^{N_2} A_{2i} + f_3 \sum_{i=1}^{N_3} A_{3i} + \ldots\ldots + f_n \sum_{i=1}^{N_n} A_{ni}}{\left(f_1 \sum_{i=1}^{N_1} A_{1i} + f_2 \sum_{i=1}^{N_2} A_{2i} + f_3 \sum_{i=1}^{N_3} A_{3i} + \ldots\ldots + f_n \sum_{i=1}^{N_n} A_{ni}\right) + f_m A_m} \quad (1)$$

where $N_1$ is the number of places under user existence probability factor $f_1$, $N_2$ is the number of places under user existence probability factor $f_2$, $N_3$ is the number of places under user existence probability factor $f_3$, $N_n$ is the number of places under user existence probability factor $f_n$.

The probability of the BS to be turned off with the number of user can be expressed as

$$p = \left(f_p\right)^n \quad (2)$$

where $f_p$ is the average importance factor considered for the entire region and n is the number of user.

The analysis is performed assuming only 15 FAPs in the considered total area. The performance would be better if more number of FAPs are used. However, FAPs should be installed in the places where there is a big possibility of users to exist. Unplanned deployment of FAPs in the network may result in no significant advantage for saving energy. Table II summarizes the values of the parameters that we used in our analysis.

TABLE I: SUMMARY OF NUMBER OF FAP AND DIFFERENT FACTORS USED IN ANALYSIS

| Station Name | No. of FAP | Factor (f) |
|---|---|---|
| Office | 3 | 1 (=$f_1$) |
| Super shop | 2 | 0.8 (=$f_2$) |
| Community center | 3 | 0.8 (=$f_2$) |
| Residence 1 | 1 | 0.7 (=$f_3$) |
| Residence 2 | 1 | 0.7 (=$f_3$) |
| Residence 3 | 1 | 0.7 (=$f_3$) |
| Residence 4 | 1 | 0.7 (=$f_3$) |
| Bank | 1 | 1 (=$f_1$) |
| Hotel | 1 | 0.7 (=$f_3$) |
| Shop 1 | 1 | 0.7 (=$f_3$) |
| Shop 2 | 0 | 0.3 (=$f_3$) |
| Remaining free space | 0 | 0.01 (=$f_3$) |

TABLE II: SUMMARY OF THE PARAMETER VALUES USED IN ANALYSIS

| Parameter | Value |
|---|---|
| Carrier frequency | 1800 [MHz] |
| Transmit signal power by the macrocellular BS | 1.5 [kW] |
| Transmitted signal power by a FAP | 15 [mW] |
| Power required for operation of a macrocellular BS | 2 [kW] |
| Power required for operation of a macrocellular BS | 8 [W] |
| Height of a macrocellular BS | 100 [m] |
| Height of a FAP | 2 [m] |
| Distance of the MS from reference FAP | 7 [m] |
| $L_{pen}$ | 20 [dB] |
| $L_{sh}$ | 8 [dB] |
| Range of a macrocellular BS | 500 [m] |
| Range of a FAP | 15 [m] |
| Average importance factor, $f_p$ | 0.7 |
| Total received noise | $7 * 10^{-7}$ [mW] |

The probability of a macrocellular BS to be turned off increases rapidly with the increase of FAPs installed in the considered network. As FAPs are very low power consuming devices, installment of a large number of FAP in the network will still result in less energy consumption compared to that by the BS alone. Fig. 5 shows the comparison of the probability for 3 different number of users existing in the network. As the number of user increases, the probability of the BS to be turned off decreases. So, the BS needs to be turned on when the number of user is more in the network as there will be a probability of at least one user to stay outside the femtocell network region.

Fig. 6 shows the comparison between 3 different cases for BS turn off probability. The energy saving possibility is enhanced if more FAPs are used. If only 5 FAPs are used in the network, the probability of the BS to be turned off is nearly 5% when 5 users exist in the network. But if 25 FAPs are used for the same network, the probability of the BS to be turned off is more than 50% for the same number of user.

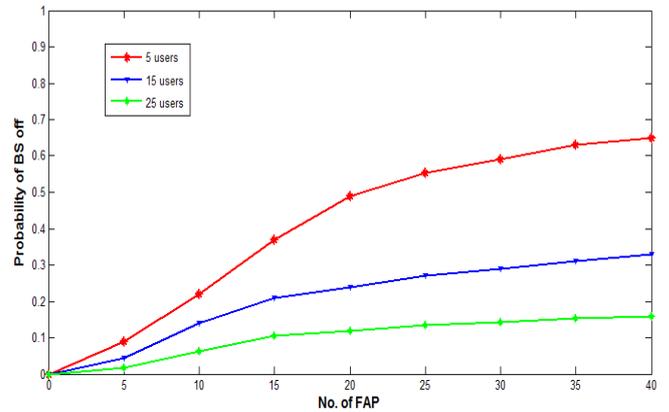

Figure 5. Probability of macrocellular BS to be turned off with increasing FAP for different number of users.

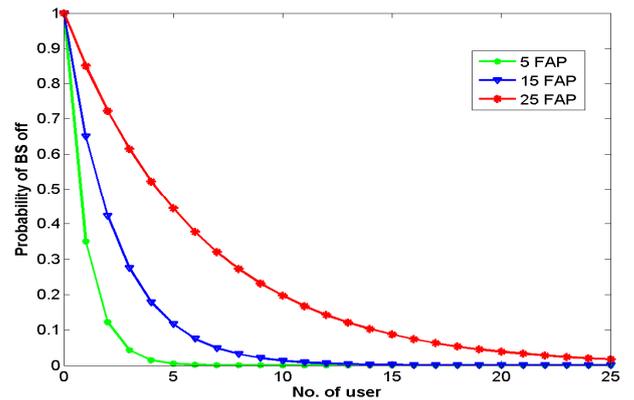

Figure 6. Comparison of macrocellular BS turn off probability with number of user for different cases.

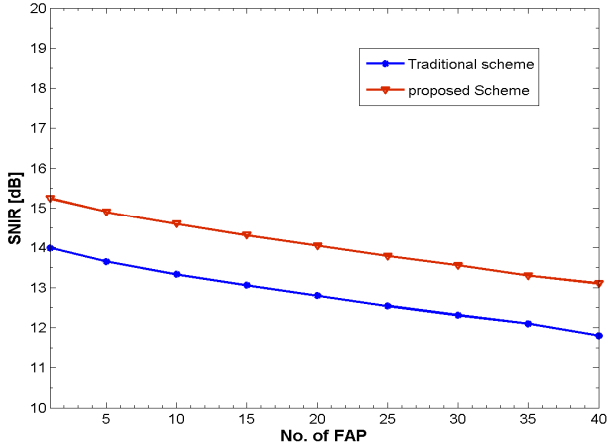

Figure 7. Comparison of SNIR levels between the proposed model and the traditional model

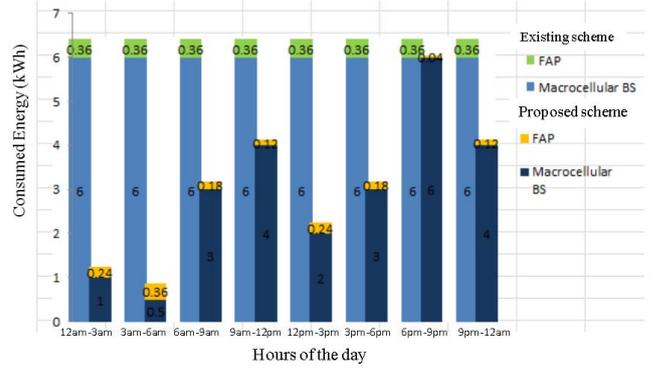

Figure 8. Comparison of energy consumption between the proposed model and traditional model

## B. SNIR Analysis

The propagation model for a macrocell case can be expressed as:

$$L = 36.55 + 26.16\log_{10} f_c - 3.82\log_{10} h_b - a(h_m) + [44.9 - 6.55\log_{10} h_m]\log_{10} d + L_{sh} \quad [dB]$$

$$a(h_m) = 1.1[\log_{10} f_c - 0.7]h_m - (1.56\log_{10} f_c - 0.8) \quad (3)$$

where $f_c$ is the center frequency in MHZ, $h_b$ is the height if the macrocellular BS in meter, $h_m$ is the height of the MS in meter, $d$ is the distance between the macrocellular BS and the MS in kilometer, $L_{sh}$ is the shadowing standard deviation.

The propagation model for a femtocell case can be expressed as [3]:

$$L_f = 20\log_{10} f_c + N\log_{10} d_1 + 4n^2 - 28 \quad [dB] \quad (4)$$

where $f_c$ is the center frequency in MHZ of the femtocell, $n$ is the number of walls between the MS and the FAP, and $d_1$ is the distance between the FAP and the MS in meter.

The received SNIR level of a femtocell user in a macrocell/femtocell integrated network can be expressed as:

$$SNIR = \frac{S_0}{I_m + I_f + N} \quad (5)$$

Simulation results show that the proposed model provides better signal-to-noise plus interference level. Fig. 7 demonstrates the comparison of SNIR level between traditional scheme and proposed scheme. The effect of the macrocellular BS and the number of FAPs has been considered here. As the BS will not remain turned on for all the time in the proposed model, there will be a decrease in the average noise level. Thus, the SNIR level is increased in the proposed model according to equation (5).

## C. Comparison of Energy Consumption

In our proposed model, the macrocellular BS will only remain shut down if the existing FAP can provide proper service to all the active users. In general, the number of active user becomes low in the middle of the night. So the BS may remain shut down for most of the time during late night. Thus, a large amount of energy can be saved in that period of night. However, the BS may remain turned off even in different times of the day if the FAPs can provide proper service to the existing users in that period of time. The FAPs consume very low energy compared to the macrocellular BS. So using many FAPs can still result in significant energy saving.

If we compare the energy consumption of the cellular network for the traditional model and the proposed model, there is a large amount of energy saved in the proposed model as the main energy consuming element for cellular network i.e., the macrocellular BS remains off for a certain period of time in this model. Fig. 8 shows the comparison of energy consumed between the traditional model and the proposed model. The saving of energy can be increased further by installation of FAPs in required places where there is a great possibility of carrying mobile users. For vehicular environment, the proposed model in [4] is good enough to serve the mobile users.

## IV. CONCLUSIONS

Saving of energy in cellular network communication is an important issue as energy prices are increasing day by day. Energy efficiency of the cellular network can be enhanced by the use of femto-access-points. Proper utilization of FAPs can result in improved service quality and reduction in the consumed energy of macrocellular base station. If the base station is allowed to turn off even for a small amount of time, a significant amount of energy can be saved as most of the energy required in the cellular communication is consumed by the base station. Our proposed model can be used extensively for energy saving purpose in cellular communications. Femtocells are a novel wireless networking technology that has several advantages including lower cost, better signal quality, and reduced infrastructure cost. In this paper, we discussed the advantages of femtocellular network for saving energy of cellular communication networks. The various performance analysis and the comparison results are presented. The simulation results demonstrate that the deployment of femtocells will significantly reduce the consumption of energy of cellular networks.